\documentclass[aps,prl,amssymb,showpacs,floatfix,twocolumn,reprint,nofootinbib]{revtex4-1}
\usepackage{amsmath,amsfonts,graphics,epsfig,amssymb,color}

\begin{document}
\title{Exact solution of the Schwarzian theory}

\author{Vladimir V. Belokurov}

\email{vvbelokurov@yandex.ru}

\affiliation{Lomonosov Moscow State University, Leninskie gory 1, Moscow, 119991, Russia and Institute for Nuclear
Research of the Russian Academy of Sciences, 60th October Anniversary
Prospect 7a, Moscow, 117312, Russia}

\author{Evgeniy T. Shavgulidze}

\email{shavgulidze@bk.ru}

\affiliation{Lomonosov Moscow State University, Leninskie gory 1, Moscow, 119991, Russia}



\begin{center}
\begin{abstract}
The explicit evaluation of the partition function in the Schwarzian theory is presented.
\end{abstract}
\end{center}
\maketitle

\vspace{1cm}

The Schwarzian theory \cite{(SW)} is the basic element of various physical models including the SYK model and the two-dimensional dilaton gravity (see, e.g., \cite{(MS)},  \cite{(BAK)}, \cite{(GR)}, \cite{(MNW)}, and references therein). The action of the theory is
\begin{equation}
   \label{Act1}
   I=-\frac{1}{g^{2}}\int \limits _{0}^{2\pi}\,\left[ \mathcal{S}_{\phi}(\tau)+\frac{1}{2}(\phi')^{2}(\tau)\right]d\tau\,.
\end{equation}
Here,
\begin{equation}
   \label{Der}
\mathcal{S}_{\phi}(t)\equiv
\left(\frac{\phi''(t)}{\phi'(t)}\right)'
-\frac{1}{2}\left(\frac{\phi''(t)}{\phi'(t)}\right)^2
\end{equation}
is the Schwarzian derivative, $\phi\in Diff^{3}([0,\, 2\pi]),\ $ and $\phi'(0)=\phi'(2\pi)\,.$

It is convenient to rewrite the action in the form
\begin{equation}
   \label{Act2}
   I=-\frac{1}{\sigma^{2}}\int \limits _{0}^{1}\,\left[ \mathcal{S}_{\varphi}(t)+2\pi^{2}\dot{\varphi}^{2}(t)\right]dt\,,
\end{equation}
where
$$
\sigma=\sqrt{2\pi}g\,,\  \  t=\frac{1}{2\pi}\tau\,,\ \ \varphi(t)=\frac{1}{2\pi}\phi(\tau) \,,
 $$
 $$
  \varphi\in Diff^{3}([0,\, 1])\,, \ \ \dot{\varphi}(0)=\dot{\varphi}(1)\, .
$$

The functional integral for the partition function
\begin{equation}
   \label{PF1}
   Z(g)=\int \limits _{\dot{\varphi}(0)=\dot{\varphi}(1)}\,\exp\left\{-I\right\}\  d\varphi
\end{equation}
diverges \cite{(SW)}. However, as we will see later on (eq. (\ref{RegPF2})), the integral
$$
 Z_{\alpha}(g)
$$
$$
=\int \limits _{  Diff^{3}([0, 1]) }
\exp\left\{-I\right\}\,\exp \left\{\frac{-2\left[ \pi^{2}-\alpha^{2}\right]}{\sigma^{2}}\int \limits _{0}^{1}\dot{\varphi}^{2}(t)dt  \right\}  d\varphi
$$
\begin{equation}
   \label{RegPF}
=\int \limits _{\dot{\varphi}(0)=\dot{\varphi}(1)}\,\exp\left\{\frac{1}{\sigma^{2}}\int \limits _{0}^{1}\,\left[ \mathcal{S}_{\varphi}(t)+2\alpha^{2}\dot{\varphi}^{2}(t)\right]dt  \right\}  d\varphi
\end{equation}
converges for $0\leq \alpha < \pi \,. $
Therefore, let us evaluate the integral (\ref{RegPF}) first.

The measure
\begin{equation}
   \label{Measure}
   \mu_{\sigma}(X)=\int \limits _{X}\,\exp\left\{\frac{1}{\sigma^{2}}\int \limits _{0}^{1}\, \mathcal{S}_{\varphi}(t)\,dt  \right\}  d\varphi
\end{equation}
is quasi-invariant, and the Radon - Nikodym derivative of the measure  is \cite{(Sh)}, \cite{(BSh)}
$$
\frac{d\mu_{\sigma}^{f}}{d\mu_{\sigma}}(\varphi)=\frac{1}{\sqrt{\dot{f}(0)\dot{f}(1)}}
$$
\begin{equation}
   \label{RadNik}
  \times \exp\left\{ \frac{1}{\sigma^{2}}\left[   \frac{\ddot{f}(0)}{\dot{f}(0)}\dot{\varphi}(0)-  \frac{\ddot{f}(1)}{\dot{f}(1)}\dot{\varphi}(1)\right]   +        \frac{1}{\sigma^{2}}\int \limits _{0}^{1}\, \mathcal{S}_{f}\left(\varphi(t)\right)\,\dot{\varphi}^{2}\,dt  \right\} \,,
\end{equation}
where
$$
\mu_{\sigma}^{f}(X)=\mu_{\sigma}(f\circ X)\,.
$$
Here, we have used the well known property of the Schwarzian derivative:
$$
\mathcal{S}_{f\circ \varphi}(t)= \mathcal{S}_{f}(\varphi(t))\dot{\varphi}^{2}(t)+ \mathcal{S}_{\varphi}(t)\,,\ \ \ \ (\,f\circ \varphi\,)(t)=f\left( \varphi(t)\right)\,.
$$
Take the function $f$ to be
\begin{equation}
   \label{f}
f(t)=\frac{1}{2}\left[ \frac{1}{\tan\frac{\alpha}{2}}\tan\left(\alpha(t-\frac{1}{2}) \right)+1 \right]\,.
\end{equation}
In this case,
$$
\mathcal{S}_{f}(t)=2\alpha^{2}\,, \ \ \ \dot{f}(0)=\dot{f}(1)=\frac{\alpha}{\sin \alpha}\,,
$$
\begin{equation}
   \label{tan}
\ \ \ -\frac{\ddot{f}(0)}{\dot{f}(0)}=\frac{\ddot{f}(1)}{\dot{f}(1)}=2\alpha\tan\frac{\alpha}{2}\,.
\end{equation}
Now we have the following functional integrals equality:
$$
\frac{\alpha}{\sin \alpha}\,\int \limits _{\dot{\varphi}(0) =\dot{\varphi}(1)}\,F(\varphi)\mu_{\sigma}(d\varphi)
$$
$$
 =\int \limits _{\dot{\varphi}(0)=\dot{\varphi}(1)}\,F(f(\varphi))\,\exp\left\{-\frac{4\alpha}{\sigma^{2}}\tan\frac{\alpha}{2}\,\dot{\varphi}(0)\right\}\,
$$
\begin{equation}
   \label{FIequality}
\times \exp\left\{\frac{1}{\sigma^{2}}\int \limits _{0}^{1}\,\left[ \mathcal{S}_{\varphi}(t)+2\alpha^{2}\dot{\varphi}^{2}(t)\right]dt  \right\}  d\varphi\,.
\end{equation}
The next step is the choice of the function $F\,.$ Let it be
\begin{equation}
   \label{F}
F(f(\varphi))=\exp\left\{\frac{4\alpha}{\sigma^{2}}\tan\frac{\alpha}{2}\,\dot{\varphi}(0)\right\}\,.
\end{equation}
To find $F(\varphi)$ from the previous equation, note that for $u(t)=f(\varphi(t))$
$$
\dot{\varphi}(0)=\frac{1}{\dot{f}(0)}\dot{u}(0)\,.
$$
Then
\begin{equation}
   \label{F(u)}
F(u)=\exp\left\{\frac{8\,\sin ^{2}\frac{\alpha}{2}}{\sigma^{2}}\,\dot{u}(0)\right\}\,.
\end{equation}
Thus for the regularized partition function we have
$$
 Z_{\alpha}(g)
$$
\begin{equation}
   \label{RegPF1}
=\frac{\alpha}{\sin \alpha}\,\int \limits _{\dot{\varphi}(0) =\dot{\varphi}(1)}\,\exp\left\{\frac{8\,\sin ^{2}\frac{\alpha}{2}}{\sigma^{2}}\,\dot{\varphi}(0)\right\}\,\mu_{\sigma}(d\varphi)\,.
\end{equation}
Under the substitution
\begin{equation}
   \label{subst}
 \varphi(t)=\frac{\int \limits _{0}^{t}\,\exp\{\xi(\eta)\}d\eta  }{\int \limits _{0}^{1}\,\exp\{\xi(\eta)\}d\eta }\,,
\end{equation}
the measure $\mu_{\sigma}(d\varphi)$ turns into the Wiener measure \cite{(Sh)}, \cite{(BSh)}
\begin{equation}
   \label{Wiener}
 w_{\sigma}(d\xi)=\exp\left\{ -\frac{1}{2\sigma^{2}}\int \limits _{0}^{1}\,\dot{\xi}^{2}(t)\,dt \right\}\ d\xi\,.
\end{equation}
In this case,
\begin{equation}
   \label{xi}
\xi(t)=\ln\dot{\varphi}(t)-\ln\dot{\varphi}(0)\,, \ \ \ \xi\in C([0,\, 1])\,,
\end{equation}
and $ \xi(0)=\xi(1)=0\,. $

Now $Z_{\alpha}(g)$ is written as
$$
 Z_{\alpha}(g)=\frac{\alpha}{\sin \alpha}
$$
\begin{equation}
   \label{RegPF2}
\times\int \limits _{\xi(0) =\xi(1)=0}\,\exp\left\{\frac{8\,\sin ^{2}\frac{\alpha}{2}}{\sigma^{2}}\,\frac{1}{\int \limits _{0}^{1}\,\exp\{\xi(\eta)\}d\eta }\right\}\,w_{\sigma}(d\xi)\,.
\end{equation}
The singularity at $\alpha=\pi $ is canceled out in the ratio
$$
\frac{Z_{\alpha}(g)}{Z_{\alpha}(1)}\,,
$$
 and we can remove the regularization there
\begin{equation}
   \label{Ratio}
   \frac{Z(g)}{Z(1)}=\lim \limits _{\alpha\rightarrow \pi-0}\ \ \frac{Z_{\alpha}(g)}{Z_{\alpha}(1)}\,.
\end{equation}

To evaluate the integral
\begin{equation}
   \label{Int}
  \int \limits _{\xi(0) =\xi(1)=0}\,\exp\left\{\frac{8}{\sigma^{2}}\,\frac{1}{\int \limits _{0}^{1}\,\exp\{\xi(\eta)\}d\eta }\right\}\,w_{\sigma}(d\xi)\,,
\end{equation}
we use the following equation:
$$
\int \limits _{\xi(0) =\xi(1)=0}\,\exp\left\{\frac{-2\beta^{2}}{\sigma^{2}(\beta +1)}\,\frac{1}{\int \limits _{0}^{1}\,\exp\{\xi(\eta)\}d\eta }\right\}\,w_{\sigma}(d\xi)
$$
\begin{equation}
   \label{Formula}
  =\frac{1}{\sqrt{2\pi}\sigma}\,\exp \left\{-\frac{4\left( \ln (\beta +1)\right)^{2}}{2\sigma ^{2}} \right\}\,.
\end{equation}
The proof of the more general formula will be given in the forthcoming paper (also, see \cite{(BSh)}).
For the integral (\ref{Int})
$$
(\beta +1)=-1\,.
$$

Thus the final result is
\begin{equation}
   \label{Result}
   \frac{Z(g)}{Z(1)}= \frac{\exp\{-\pi\}}{g}\,\exp\left\{ \frac{\pi}{g^{2}}\right\}\,.
\end{equation}

It is interesting to compare the one-loop results for $Z(g)$ in \cite{(SW)} with the eq. (\ref{Result}). Note that the power of the constant $g$ in the denominator is determined by the number of gauge fixing conditions. The one-loop result for the orbit $Diff(S^{1})/U(1)$ (eq. (3.45) in \cite{(SW)}) has the same form as our exact result (\ref{Result}).

Unlike its compact subgroup $U(1)$, the group $SL(2, \textbf{R})$ is noncompact. Therefore, integrating over the quotient space $Diff^{3}([0, 1])/SL(2,\textbf{R}) $
we get the finite result for the partition function in the Schwarzian theory.

We define the Schwarzian partition function as a limit
\begin{equation}
   \label{Ratio2}
  Z_{Schw}(g) = \lim \limits_{\alpha\rightarrow \pi - 0}\ \frac{Z_{\alpha}(g)}{V_{\alpha}(g)} \,.
\end{equation}
Here, $ Z_{\alpha}(g) $ is given by the eqs. (\ref{RegPF}), (\ref{RegPF2}), and $V_{\alpha}(g) $ is the regularized volume of the group $SL(2,\textbf{R})$
\begin{equation}
   \label{V}
V_{\alpha}(g)=\int \limits _{SL(2,\textbf{R})}\exp \left\{\frac{-2\left[ \pi^{2}-\alpha^{2}\right]}{\sigma^{2}}\int \limits _{0}^{1}\dot{\varphi}^{2}(t)dt  \right\}  d\mu _{H}\,.
\end{equation}
Note that the functional measure in the eq. (\ref{RegPF}) and the Haar measure $d\mu _{H}$ on the group $SL(2, \textbf{R})$  in the eq. (\ref{V}) are regularized in the same manner.

To perform the integration over the group $SL(2,\textbf{R})$ in the eq. (\ref{V}) we choose the representation \cite{(Lang)}
\begin{equation}
   \label{Repres}
  \varphi_{z}(t)=-\frac{i}{2\pi}\ln\frac{e^{i2\pi t}+z}{\bar{z}\,e^{i2\pi t}+1}\,, \ \ \ z=\rho e^{i\theta}\,, \ \ \ \rho < 1  \,.
\end{equation}
In this case, the Haar measure is \cite{(Lang)}
\begin{equation}
   \label{Haar}
  \mu_{H}(dz)=\frac{4\rho d\rho \,d\theta}{\left(1-\rho^{2} \right)^{2}} \,.
\end{equation}

The integral
\begin{equation}
   \label{dt}
 \int \limits _{0}^{1}\dot{\varphi}_{z}^{2}(t)dt =\int \limits _{0}^{1}\frac{\left(1-|z|^{2} \right)^{2} \,dt}{\left( e^{i2\pi t}+z \right)^{2}\left(e^{-i2\pi t}+\bar{z} \right)^{2}}
 =\frac{1+\rho^{2}}{1-\rho^{2}}
\end{equation}
does not depend on $\theta \,.$
And the regularized volume of the group has the form
$$
V_{\alpha}(g)
=\int \limits _{0}^{1}\exp \left\{-\frac{\left[\pi^{2}-\alpha^{2}\right]}{\pi g^{2}} \frac{\left(1+\rho^{2}\right)}{\left(1-\rho^{2}\right)} \right\}
\frac{8\pi \rho d\rho }{\left(1-\rho^{2} \right)^{2}}
$$
\begin{equation}
   \label{Valphag}
=\exp \left\{-\frac{\left[\pi^{2}-\alpha^{2}\right]}{\pi g^{2}}\right\} \frac{4\pi ^{2}g^{2}}{\left[\pi^{2}-\alpha^{2}\right]}\,.
\end{equation}

Thus we can evaluate the Schwarzian partition function
\begin{equation}
   \label{Result2}
   Z_{Schw}(g)= \frac{1}{4\pi g^{3}}\,\exp\left\{ \frac{\pi}{g^{2}}\right\}\,.
\end{equation}

Note that the one-loop result in \cite{(SW)}, \cite{(MS)} has the same form as the exact partition function (\ref{Result2}) obtained by the direct functional integration.

\end{document}